\documentclass[
aps,%
12pt,%
final,%
oneside,%
nobibnotes,%
nofootinbib,%
superscriptaddress,%
centertags,
floatfix,
secnumarabic]%
{revtex4}%

\usepackage{braket}
\usepackage[english]{babel}
\usepackage{natbib}

\usepackage[hidelinks]{hyperref}
\usepackage{xcolor}
\usepackage{multirow}
\usepackage{siunitx}
\usepackage{indentfirst}
\usepackage{graphicx}
\DeclareGraphicsExtensions{.pdf,.eps}
\usepackage{amsmath}
\usepackage{cancel}
\usepackage{booktabs}

\newcommand{\M}{\mathcal{M}}

\newcommand{\MeV}{\,\text{MeV}}
\newcommand{\GeV}{\,\text{GeV}}
\newcommand{\Br}{\mathrm{Br}}
\newcommand{\B}{\mathcal{B}}

\begin{document}
\title{Weak decays of doubly heavy baryons: decays to system of $\pi$ mesons}
\author{A.S. Gerasimov}
\email{Anton.Gerasimov@ihep.ru}
\affiliation{NRC "Kurchatov Institute", IHEP, Protvino, Moscow Region, 142281, Russia}
\affiliation{Moscow Institute of Physics and Technology}

\author{A.V. Luchinsky}
\email{alexey.luchinsky@ihep.ru}
\affiliation{NRC "Kurchatov Institute", IHEP, Protvino, Moscow Region, 142281, Russia}
\affiliation{Bowling Green State University, Bowling Green, Ohio, 43402, USA}

\begin{abstract}
In the presented article we consider exclusive weak decays of doubly heavy baryons with spin $J=1/2$ with production of leptonic pair or a set of light mesons. 
Using QCD factorization theorem and spectral functions approach we obtain theoretical predictions for the partial probabilities of these decays and distributions over various kinematic variables. According to obtained results partial probabilities of some of the considered decays are large enough to be observed experimentally.
\end{abstract}
  
\maketitle

\section{Introduction}
Today the research of doubly heavy baryons takes a particularly important role in physics of elementary particles. For a long time they were not available for experimental study. In 2005 SELEX collaboration has reported \cite{Ocherashvili:2004hi} the observation of $\Xi_{cc}^{+}$ baryon in the decay $\Xi_{cc}^{+}\to p D^{+}K^{-}$. This result was not confirmed, but later the LHCb collaboration managed to register doubly heavy baryon $\Xi_{cc}^{++}$ in $\Lambda^+_c K^- \pi^+ \pi^+$ final state \cite{Aaij:2017ueg}. Thereby we are interested in theoretical research of other decays of doubly heavy baryons for subsequent experimental verification.

In this article decays of ground state doubly heavy baryons are considered: $\B_1 \to \B_2 R$, where $R=\mu\nu_{\mu}$, $\pi$, $2\pi$, $3\pi$, $5\pi$. According to the factorization theorem they are related to $\tau$-lepton decays $\tau \to \nu_{\tau} R$, since in both cases the system $R$ is produced by virtual $W$-boson transition $W\to R$. As a result probability of reaction $\B_{1}\to\B_{2}W$ is defined by convolution of semi-leptonic differential width $\B_{1}\to\B_{2}\mu\nu$ and spectral function of transition $W\to R$. First process can be described in terms of form factors of weak decay, which can be calculated with help of potential models. Spectral functions, on the other hand, can be found by analyzing decays of $\tau$-lepton $\tau\to\nu_\tau R$.

The rest of the paper is organized as follows. In the next section description of used in our paper theoretical model and parametrization of the form factor are given (see also appendix \ref{app2}). Section \ref{sec:spectral} is devoted to description of the spectral functions. In the next section  we give our numerical results: integrated branching fractions  of the considered decays and squared momentum distribution. Discussion of the  obtained results is given in the final section of the article.

\section{Theoretical Description}
\label{sec:Th}

Let us consider the reaction
\begin{align}
\B_1 \to \B_2 \  R,
\end{align}
where doubly heavy baryon $\B_1$ with spin $J=1/2$ decays into another doubly heavy baryon $\B_2$ with same spin and virtual $W$-boson, which then hadronizes into system $R$. where $R = \ell \nu_\ell$, $\pi$, $2\pi$, $3\pi$, $5\pi$.

\begin{figure}
\begin{minipage}[h]{0.47\linewidth}
\includegraphics[width=\textwidth]{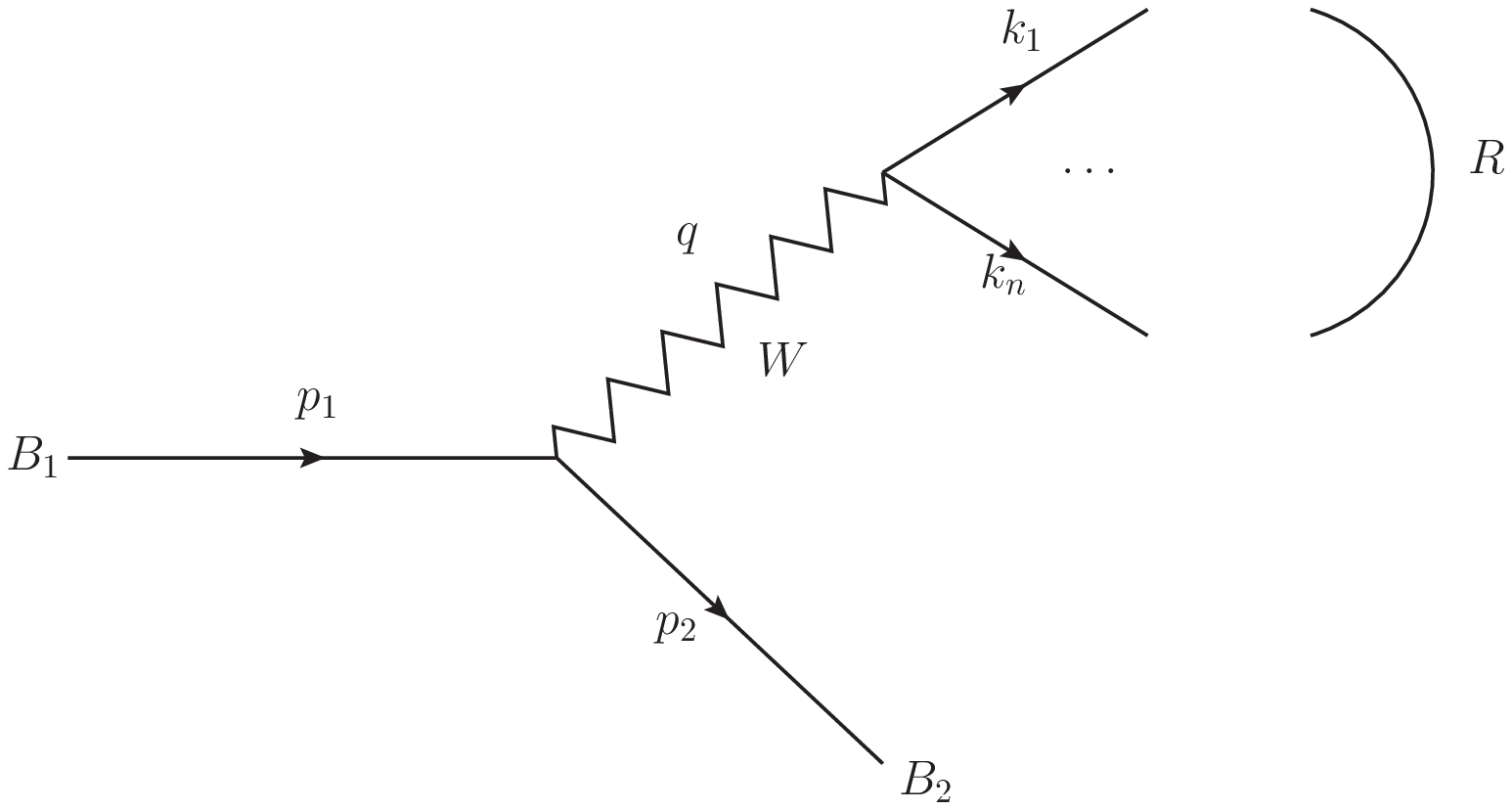}
\end{minipage}
\begin{minipage}[h]{0.47\linewidth}
\includegraphics[width=\textwidth]{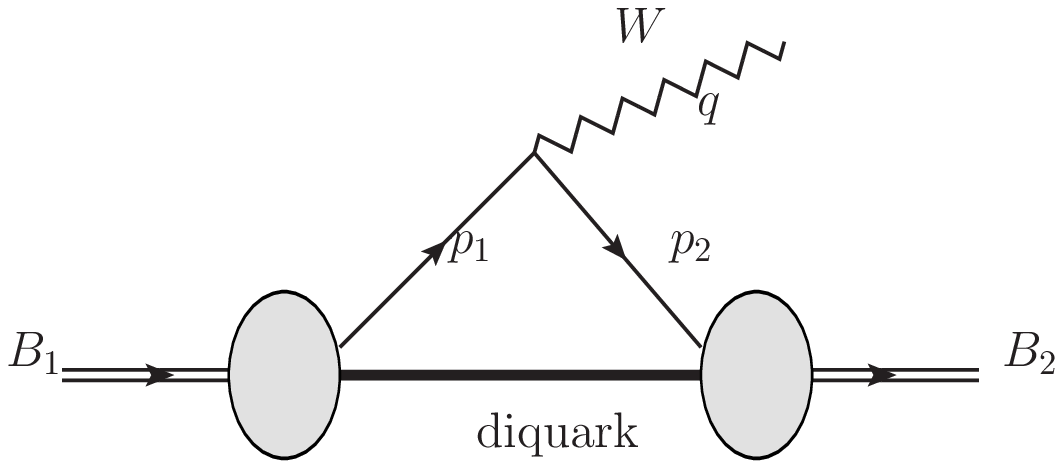}
\end{minipage}
\caption{Typical diagrams for weak decays $\B_1 \to \B_2 \ R$\label{diagramR}}

\end{figure}

The considered baryons contain heavy $b$- and $c$-quarks, so at the quark level their decays are caused by  $c \to s/d$ and $b \to u/c$ transitions. Typical diagram of such processes is shown in figure \ref{diagramR}. We should note that there exists also an unfactorized case, when light quark from baryon takes part in production of  $\pi$. However, contribution of such a diagram is small due to color suppression and its calculation is not a subject of this article.
 
According to factorization theorem process $\B_1 \to \B_2 \ R$ can be separated into two independent parts: weak decay $\B_1 \to \B_2 \ W$ and $W \to R$ transition. As a result its  matrix element can be written as
\begin{align}
  \label{eq:matr}
\M &= \frac{G_F}{\sqrt 2} V_{ij} a_1 H^\mu \varepsilon_\mu,
\end{align}
where $G_F$ is the Fermi constant, $V_{ij}$ the CKM matrix element, constant $a_1$ describes soft gluon rescattering \cite{Buchalla:1995vs}, $\varepsilon_{\mu}$ is the effective polarization vector of $W$-boson, and the amplitude of $\B_1 \to \B_2$ transition is denoted as $H^{\mu}$.

\subsection{$\B_{1}\to\B_{2}W$ Transition}
\label{sec:bbw}

This amplitude can be written as \cite{Wang:2017mqp}
\begin{align}
H^{\mu}=\overline{u}(P^{\prime})\left[\gamma^{\mu}f_{1}(q^2)+i\sigma^{\mu\nu}\frac{q_\nu}{M_1}f_{2}(q^2)+\frac{q^\mu}{M_1}f_{3}(q^2)\right]u(P)- \nonumber\\
-\overline{u}(P^{\prime})\left[\gamma^{\mu}g_{1}(q^2)+i\sigma^{\mu\nu}\frac{q_\nu}{M_1}g_{2}(q^2)+\frac{q^\mu}{M_1}g_{3}(q^2)\right]\gamma_{5}u(P),
\end{align}
where
\begin{align}
\sigma^{\mu \nu}&=i(\gamma^{\mu}\gamma^{\nu}-\gamma^{\nu}\gamma^{\mu})/2,
\end{align}
$M_1$ is the mass of the initial baryon, $P (P^{\prime}$) is the momentum of initial (final) baryon,  $q=P'-P$ is the transferred momentum, and $f_i(q^{2}), g_i(q^{2})$ are the form factors of $\B_{1}\to\B_{2}$ transition. Contribution of $f_3$ and $g_3$ form factors are suppressed as $(m_u+m_d)/m_\B$ and we will neglect them in the following.

In the calculation of form factors we use quark-diquark model, in which baryon is considered to be build from scalar or axial diquark (with spin $S_{d}=0$ and $1$ respectively) and a single quark. It is convenient to take those quarks  as a diquark, that do not take part in weak decay. It is clear that in this approximation spin and internal state of diquark remain the same and, for example, in case of $\Omega_{bc}^{0}\to\Omega_{cc}^{+}W^{-}$ decay it is required to consider form factors of
\begin{align}
 b (cs)_{S,A}\to c (cs)_{S,A}+W^{-},
\end{align}
transitions. In the above expression indexes $S$, $A$ correspond to scalar and axial diquark. As a result, the form factor of this decay is equal to
\begin{align}
    F(q^2)=c_S F_S(q^2)+c_A F_A(q^2),
\end{align}
where $F_{S,A}(q^2)$ and $c_{S, A}$ the form factors and coefficients of scalar (axial) diquarks respectively. The form factors can be calculated in the framework of potential model and, according to \cite{Wang:2017mqp}, are determined by overlap integrals. Detailed presentation of calculation of form factors can be found in \cite{Shi:2019hbf, Wang:2017mqp}. As for the coefficients $c_{S,A}$, their values can be determined using Clebsh-Gordon coefficients.

Let us consider baryon $\Omega_{cc}^+$ as an example. In this case diquark $cc$ can only have spin $S_{d}=1$ (scalar diquark option is forbidden by Pauli exclusion principle), so its spinor wave function can be written as
\begin{align}
   (c c)_A &= c_1(\uparrow) c_2(\uparrow),
\end{align}
where symbols $\uparrow$ and $\downarrow$ mark states of $c$-quark with spin projections $S_{z}=1/2$ and $-1/2$ respectively. The wave function of the baryon with full spin $S=1/2$ can be written in the same way:
\begin{align}
 \Omega_{cc}^+ &= c_1(\uparrow) c_2(\uparrow) s(\downarrow).
\end{align}
In $c(cs)$ basis, on the other hand, this wave function changes to
\begin{align}
  \Omega_{cc}^+ &= \frac{1} {\sqrt[]{2}}
                  \left[-\frac{\sqrt[]{3}}{2} c_1(\uparrow) (c_2s)_S + \frac{1}{2}c_1(\downarrow)(c_2s)_A + (c_1 \leftrightarrow c_2)\right].
\end{align}
For baryon $\Omega_{bc}^0$ the wave function is
\begin{align}
\Omega_{bc}^0=-\frac{\sqrt[]{3}}{2} b(\uparrow) (cs)_S + \frac{1}{2}b(\downarrow)(cs)_A 
\end{align}
From these expressions for decay $\Omega_{bc}^0 \to \Omega_{cc}^+\  R$ we get 
\begin{align}
c_S &= \frac{3\sqrt[]{2}}{4}, \qquad c_{A}=\frac{\sqrt[]{2}}{4}.
\end{align}
In Appendix \ref{app2} we show the values of these coefficients and parametrization of the form factors for all considered in our article decays.

\subsection{Spectral Function Formalism}
\label{sec:spectr}

The width of the $\B_{1}\to\B_{2}R$ decay is equal to 
\begin{align}
\Gamma=\frac{1}{2}\frac{1}{2M_1}\int d\Phi_{n+1} |\M|^2,
\end{align}
where Lorentz-invariant phase space $d\Phi_{n}$ is defined by the expression
\begin{align}
  d\Phi_{n}(q \to k_1 \dots k_n)&=\left(2\pi \right)^4 \delta^4\left(q-\sum k_i \right)\prod \frac{d^3k_i}{(2\pi)^3 2E_i}.
  \label{eq:LIPS}
\end{align}
The following recurrence relation holds for this phase space:
\begin{align}
d\Phi_{n+1}(p_1\to p_2\ k_1\dots k_n) &= \frac{dq^2}{2\pi}d\Phi_2(p_1\to p_2\ q)d\Phi_n(q \to k_1 \dots k_n)=\nonumber\\
 &=dq^2\frac{\lambda}{8\pi}\frac{d\Phi_n(q \to k_1 \dots k_n)}{2\pi},
\end{align}
where
\begin{align}
\lambda=\sqrt{1-\left( \frac{M_2+\sqrt{q^2}}{M_1} \right)^2}\sqrt{1-\left( \frac{M_2-\sqrt{q^2}}{M_1} \right)^2},
\end{align}
and $M_2$ is the mass of the final baryon. Summed over polarizations squared  matrix element \eqref{eq:matr} an be written in the following form:
\begin{align}
\sum|\M|^2&=\frac{{G_F}^2}{2}V_{ij}^2 a_1^2 H^{\mu}H^{\nu*} \varepsilon_{\mu}\varepsilon^*_{\nu}.
\end{align}
Decay width distribution over square transferred momentum with help of these expressions can be written as
\begin{align}
  \label{eq:dGdq2}
\frac{d \Gamma}{dq^{2}}=\frac{G_F^2}{2}V_{ij}^2a_1^2\frac{1}{2M_1}\frac{1}{2} \frac{\lambda}{8\pi} \left(H_T^2\rho_T+H_L^2\rho_L\right),
\end{align}
where $\rho_{L,T}(q^{2})$ are (dependent on the final state $R$) longitudinal and transverse spectral functions, defined according to
\begin{align}
\int \frac{d\Phi_{n}(q \to k_1 \dots k_n)}{2\pi}
\varepsilon_{\mu}\varepsilon^{*}_{\nu}
=(q_{\mu}q_{\nu}-q^2g_{\mu \nu})\rho_T(q^2)+q_{\mu}q_{\nu}\rho_L(q^2),
\end{align}
while squared longitudinal and transverse matrix elements are equal to
\begin{align}
H_T^2&=H^{\mu}H^{\nu*}(q_{\mu}q_{\nu}-q^2g_{\mu \nu})=
\frac{1}{2M_1^2}(f_1^2 M_1^2 (-2 q^4+2 q^2 M_{-}^2-q^2 M_{+}^2+M_{-}^2 M_{+}^2)+
\nonumber\\ &
12 f_1 f_2 q^2 M_{+} (q^2-M_{-}^2) M_1-4 f_2^2 q^2 (q^4-q^2 M_{-}^2+2 q^2 M_{+}^2-2 M_{-}^2 M_{+}^2)+
\nonumber\\ &
g_1^2 (M_{-}+M_{+})^2 (-2 q^4-q^2 M_{-}^2+2 q^2 M_{+}^2+M_{-}^2 M_{+}^2)+12 g_1 g_2 q^2 M_{-} (M_{+}^2-q^2) (M_{-}+M_{+})-
\nonumber\\ &
4 g_2^2 q^2 (q^4+2 q^2 M_{-}^2-q^2 M_{+}^2-2 M_{-}^2 M_{+}^2)) \nonumber
\\ 
H_L^2&=H^{\mu}H^{\nu*}q_{\mu}q_{\nu}= 2 \bigg( f_1^2 M_{-}^2
   \left(M_{+}^2-\text{q}^2\right)+g_1^2 M_{+}^2 \left(M_{-}-\text{q}^2\right) \bigg). 
\end{align}
where $M_\pm=M_1\pm M_2$.

\section{Calculations of spectral functions}
\label{sec:spectral}
\begin{figure}
\center{\includegraphics[width=0.6\textwidth]{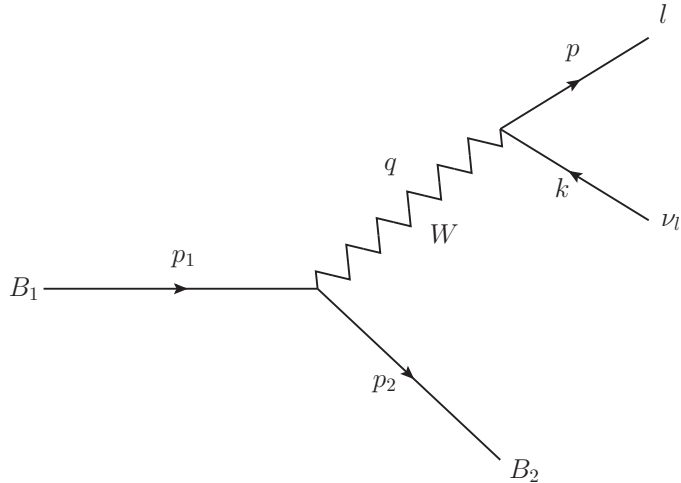}}
\caption{Diagram for semi-leptonic decay \label{fig:lnul}}
\end{figure}

In several cases the spectral functions can be easily calculated. For example, for semi-leptonic decay (see its diagram on pic. \ref{fig:lnul}) effective polarization vector of $W$-boson equals to
\begin{align}
\varepsilon_{\mu}=\overline{u}_{\nu_l}(k)\gamma_{\mu}(1+\gamma_5)u_l(p),
\end{align}
where $p$, $k$ are the momentum of final leptons (we assume that their masses are negligible). It is easy to show that summed over leptons' polarization square of this vector is equal to
\begin{align}
  \sum_{\mathrm{pol}}\varepsilon_{\mu}\varepsilon^{*}_{\nu}=8 \left( -g_{\mu \nu}(p k)+k_{\mu}p_{\nu}+k_{\nu}p_{\mu}+i e_{\mu\nu\alpha\beta}p^\alpha k^\beta\right).
  \label{eq:EpsMN}
\end{align}
As a result in this case longitudinal spectral function identically equal to zero (we could expect this because of partial conservation of the axial current). For the transversal one we have 
\begin{align}
\rho_T^{(l \nu_l)}(q^2)=\frac{1}{6\pi^2}.
\end{align}

Diagrams for decays to $\pi$ and $\rho$ mesons are trivial. It is well known that amplitudes of $W \to \pi$ and $W \to \rho$ transitions are defined as
\begin{align}
\braket{\pi|J_{\mu}|W}&=f_{\pi}k_{\mu},\quad
\braket{\rho|J_{\mu}|W}=f_{\rho}m_{\rho}\varepsilon_{\mu}^*,
\end{align}
respectively, where $f_{\pi}=130\MeV$, $f_{\rho}=216\MeV$, $m_{\rho}=775\MeV$ \cite{Tanabashi:2018oca}.
As a result we have:
\begin{align}
\rho_T^{(\pi)}(q^2)&=0, &&\rho_L^{(\pi)}=f_{\pi}^2\delta(q^{2}-m_{\pi}^{2}),\\
\rho_T^{(\rho)}(q^2)&=f_{\rho}^2\delta(q^{2}-m_{\rho}^{2}), &&\rho_L^{(\rho)}=0.
\end{align}

\begin{figure}
\center{\includegraphics[width=0.9\textwidth]{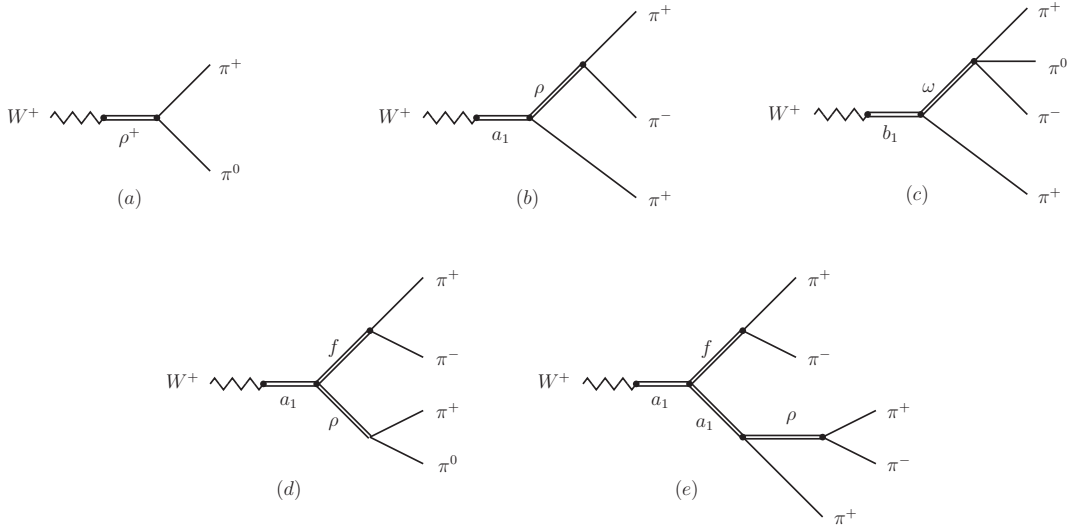}}
\caption{Diagrams for transitions $W\to 2\pi$, $W\to 3\pi$, $W\to 4\pi$ and $W\to 5\pi$\label{fig:diagsWPi}}
\end{figure}

For $W\to 3\pi$, $4\pi$, $5\pi$ transitions the longitudinal spectral functions are equal to zero because of partial conservation of the axial current. The shapes of  the transversal functions, on the other hand, can be determined using resonance model (see shown in Fig.\ref{fig:diagsWPi} diagrams). The normalization of these functions can be found using experimental values of the branching fractions of $\tau\to\nu_{\tau}+R$ decays \cite{Tanabashi:2018oca, PhysRevLett.90.181802, Likhoded:2013iua}. For these processes 
\begin{align}
H^{\mu}&=\overline{u}(P_{\nu_{\tau}})\gamma^{\mu}(1+\gamma_{5})u)(P_{\tau})\\
\Gamma({\tau \to \nu_{\tau} R})&=\frac{G_F^2}{2}\frac{1}{2m_{\tau}} \frac{\lambda}{8\pi} \int dq^2 \Big(m_{\tau}^4  +m_{\tau}^2 q^2 -2q^4 \Big)\rho_T,
\end{align}
where $m_{\tau}=1776\MeV$ \cite{Tanabashi:2018oca} is the mass of $\tau$ lepton, $P_{\tau}$ is its momentum, and $P_{\nu_{\tau}}$ is the  momentum of $\tau$-neutrino.
Analytical calculations of the corresponding spectral functions is rather complicated, so we made a numerical integration using created by our group EvtGen models \cite{Lange:2001uf}.

\begin{figure}
  \centering
  \includegraphics[width=0.8\textwidth]{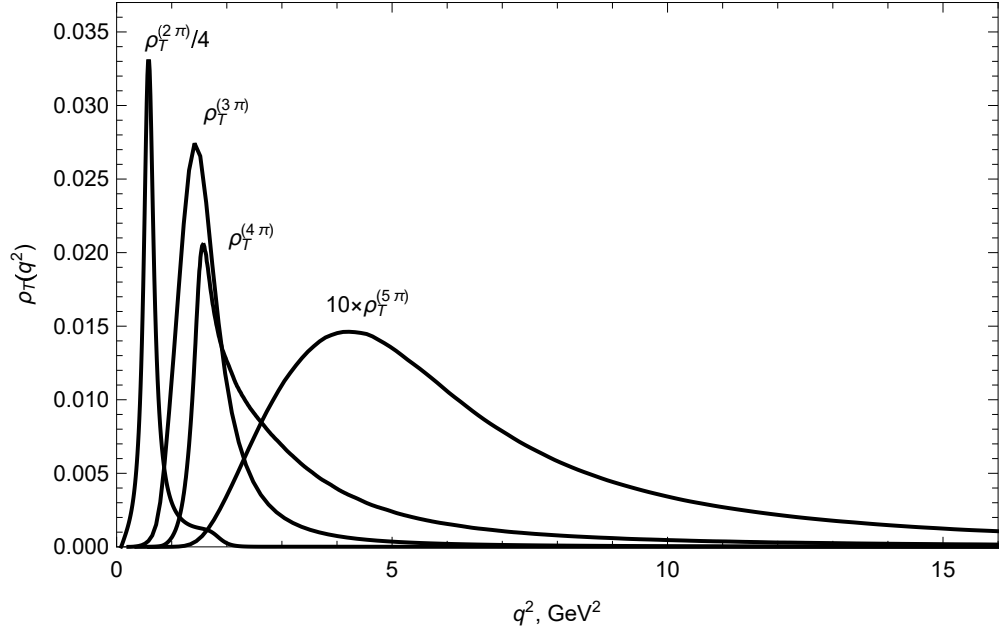}
  \caption{Spectral functions}
  \label{fig:rhoT}
\end{figure}

In figure \ref{fig:rhoT} we show $q^{2}$ dependence of these spectral functions.
It can be easily seen that the $\rho_{T}^{2\pi}$ spectral function has a clear peak in $q^{2}\approx 0.5\,\GeV^{2}$ region and a little bump at somewhat higher energy. These peaks correspond to contributions of virtual $\rho$ meson and its excitations. As for $\rho_{T}^{(3\pi)}$, this spectral function has a peak in $q^{2}\approx 1\GeV^{2}$ region. This peak corresponds to shown in diagram \ref{fig:diagsWPi}(b) virtual $a_{1}$ meson, whose mass is $m_{a_{1}}\approx 1.2\,\GeV$ \cite{Tanabashi:2018oca}. The same meson also produces peaks $\rho_{T}^{(4\pi)}$ and $\rho_{T}^{(5\pi)}$ spectral functions, but they are shifted to higher energy region by multi-body phase space.
It is interesting to note also that in the case of $4\pi$ production both $a_{1}$ and $b_{1}$ mesons can contribute (see Figs. \ref{fig:diagsWPi}(c) and (d)). Theoretically we should consider the interference between these two  channels, but, since in the latter case three pions are produced in  the decay of the narrow $\omega$ meson, such interference can be  neglected.

Comparison with later experimental results \cite{Aaij:2016rks,Khachatryan:2014nfa,Aaij:2014bla,LHCb:2012ag} showed that these spectral functions suits well for theoretical description of exclusive decays of $B_{c}$-mesons, so we can expect that in case of studied in our article processes they can also be used.

\section{Numerical results}
\label{sec:num}

For numerical calculation we use values for Fermi constant and CKM matrix elements given in PDG \cite{Tanabashi:2018oca}. 
\begin{align}
G_F&=1.166\times 10^{-5}\,\GeV^{-2}, \nonumber \\
V_{ud}&=0.974, \qquad V_{us}=0.225, \qquad V_{ub}=0.00357, \nonumber \\
V_{cd}&=0.225, \qquad V_{cs}=0.974, \qquad V_{cb}=0.0411.
\end{align}
Values of masses of initial and final baryons are presented in table \ref{tab:mass_width}.

\begin{table}
  \centering
 \begin{tabular}{@{}llllll||llllll@{}}
    \hline
  $\B$ & $m_{\B}$ & $\Gamma_{\B}$ &   $\B$ & $m_{\B}$ & $\Gamma_{\B}$ &
         $\B$ & $m_{\B}$ & $\B$ & $m_{\B}$  &$\B$ & $m_{\B}$ \\
  \hline
    $\Xi_{cc}^{++}$ & $3.627$ & $3.3$  & $\Xi_{cc}^{+}$ & $3.627$ & $10.$ &
         $\Lambda_{c}^{+}$ & $2.286$  & $\Sigma^{++}_{c}$ & $2.454$ &
         $\Sigma^{+}_{c}$ & $2.453$\\
    $\Omega_{cc}^{+}$ & $3.65$ & $3.7$  & $\Xi_{bb}^{0}$ & $10.31$ & $2.7$  &
         $\Sigma^{0}_{c}$ & $2.454$ &
         $\Xi_{c}^{+}$ & $2.486$ & 
         $\Xi_{c}^{\prime +}$ & $2.576$ \\
$\Xi_{bb}^{-}$ & $10.32$ & $2.7$  & $\Omega_{bb}^{-}$ & $10.45$ & $1.2$  &
         $\Xi_{c}^{0}$ & $2.471$ & 
         $\Xi_{c}^{\prime 0}$ & $2.578$ &
         $\Omega_{c}^{0}$ & $2.695$ \\
$\Xi_{bc}^{+}$ & $6.914$ & $4.1$  & $\Xi_{bc}^{0}$ & $6.914$ & $11.$ &
 $\Lambda_{b}^{0}$ & $5.620$ &
 $\Sigma_{b}^{+}$ & $5.811$ & $\Sigma_{b}^{0}$ & $5.814$\\
$\Omega_{bc}^{0}$ & $7.136$ & $4.5$  & $\Xi_{bc}^{\prime+}$ & $6.914$ & $4.1$ &
         $\Sigma_{b}^{-}$ & $5.816$ & 
         $\Xi_{b}^{0}$ & $5.793$ &
         $\Xi_{b}^{\prime 0}$ & $5.935$ \\
$\Xi_{bc}^{\prime0}$ & $6.914$ & $11.$  & $\Omega_{bc}^{\prime0}$ &  $7.136$ & $4.5$  &
         $\Xi_{b}^{-}$ & $5.795$ &
         $\Xi_{b}^{\prime -}$ & $5.935$ & $\Omega_{b}^{-}$ & $6.046$ \\
  \hline
  \end{tabular}  \caption{Masses (in GeV) )and decay widths (in $\textrm{ps}^{-1}$) of initial and final doubly heavy baryons}
  \label{tab:mass_width}
\end{table}

Using presented in the previous section expressions it is easy to calculate the branching fractions of the considered in our article decays (see tables \ref{tab:resCC}, \ref{tab:resBC}, \ref{tab:resBB}).
The branching fraction of doubly heavy baryons decays with semileptonic pair, $\pi$, or $\rho$ mesons production agree perfectly with the results of paper \cite{Wang:2017mqp}. It should be stressed, however, that in the case of the latter final state the width of the $\rho$ meson is completely neglected in this work. It is clear, however, that in some cases such approximation could be not reliable. As a result, obtained in the framework of spectral functions formalism branching fractions (labelled $2\pi$ in tables \ref{tab:resCC}, \ref{tab:resBC}, \ref{tab:resBB}) differ from the results with $\rho$-meson width neglected (see columns labeled $\rho$ in these tables).  It should be noted also that depending on the initial and final baryons the excitations of the $\rho$ meson increase the branching fraction of the $\B_{1}\to\B_{2}2\pi$ decays
 by approximately $1\div 10\%$).

\begin{figure}
  \centering
  \includegraphics[width=0.45\textwidth]{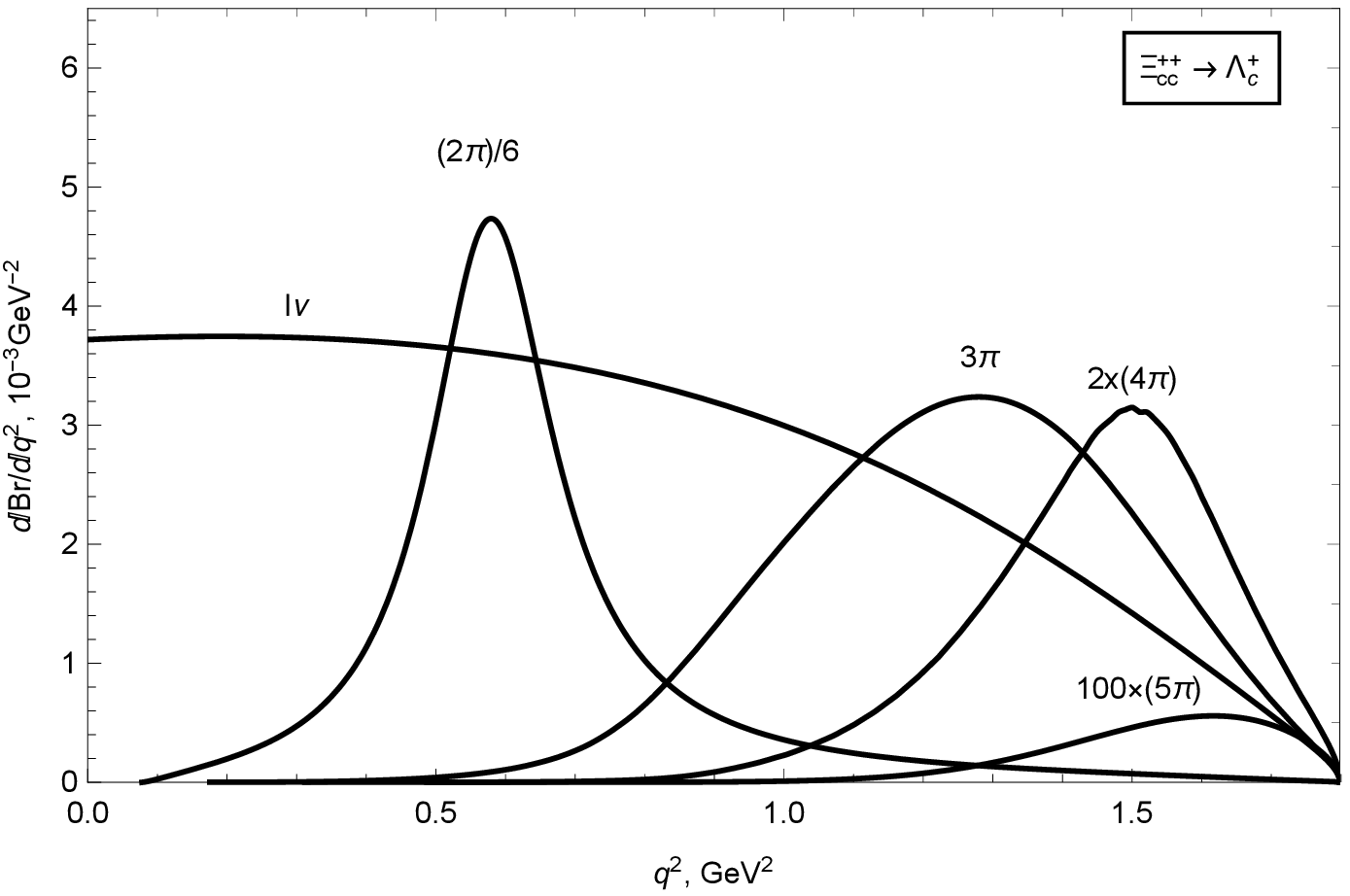}\includegraphics[width=0.45\textwidth]{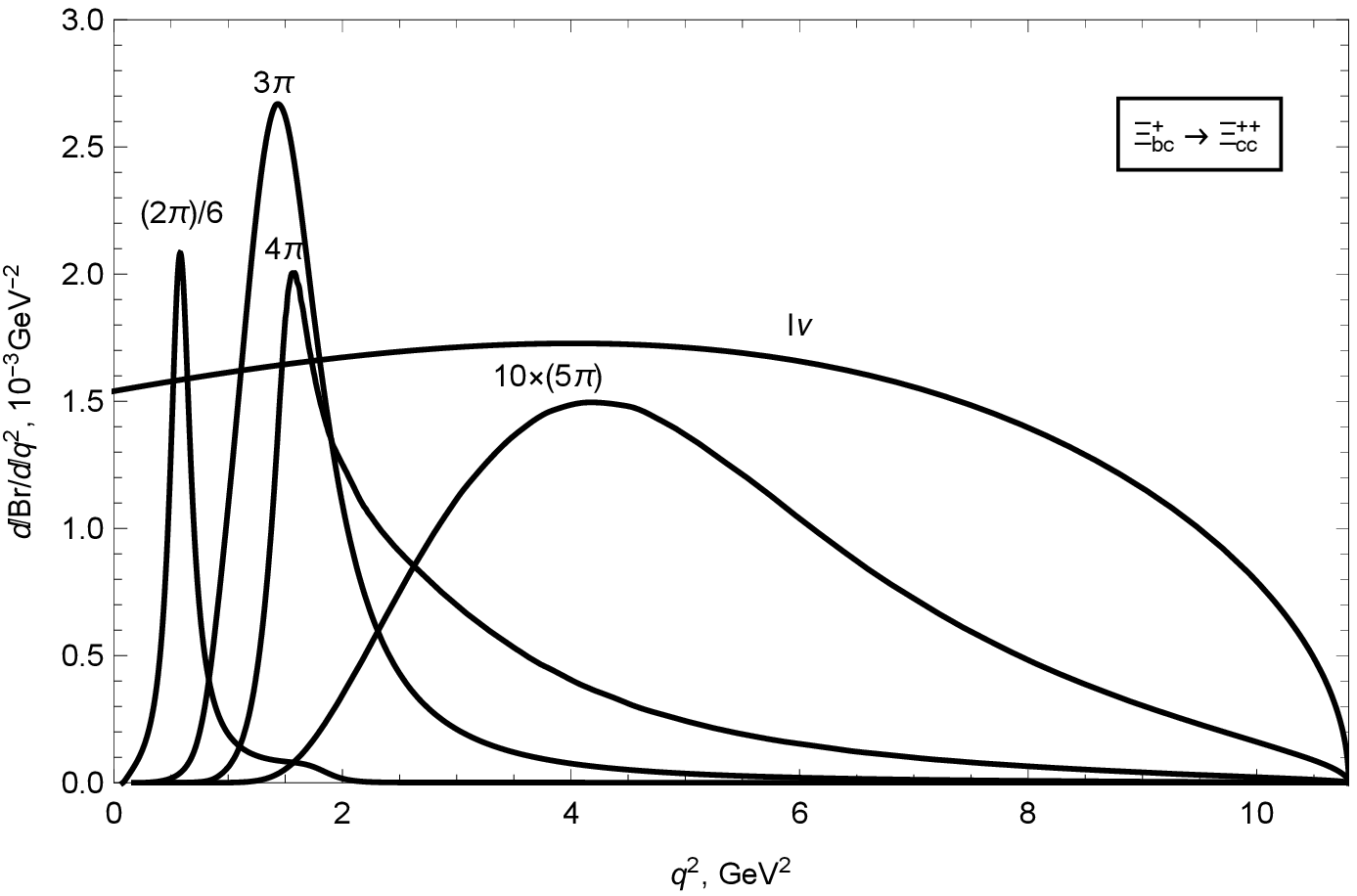}
  \caption{$q^2$ distributions of the $\Xi_{cc}^{++}\to\Lambda_{c}^{+}$ and $\Xi_{bc}^{+}\to\Xi_{cc}^{++}R$ branching fractions (left and right figures respectively) for $R=\ell\nu$, $2\pi$, $3\pi$, $4\pi$, $5\pi$. Note that some curves are scaled as shown in the figures.}
  \label{fig:q2_npi}
\end{figure}

The branching fractions of $3\pi$, $4\pi$ and $5\pi$ production are given in the last three columns of these tables. One can easily see, that typically these processes are suppressed in comparison, for example, with single $\pi$ in the final state. It should be noted, however, that in the case of $b$ quark decay  (see $\Xi_{bc}^{+}\to\Xi_{cc}^{++}$, for example) such suppression is not observed. The reason is that in this case the energy deposit is large enough, so higher $q^2$ values are reachable. This is exactly the region where, according to fig. \ref{fig:rhoT},  spectral functions of $W \to 4\pi$, $5\pi$ transitions are large. In the case of c-quark decay, on the other hand, only low $q^2$ region is available kinematically and in this region spectral functions of multiple quark production are small. This effect can be clearly seen in figure \ref{fig:q2_npi}, where all $q^{2}$ distributions are shown both for $\Xi_{cc}^{++}\to\Lambda_{c}^{+}$ and $\Xi_{bc}^{+}\to\Xi_{cc}^{++}$ transitions.

\begin{figure}
\center{\includegraphics[width=0.9\textwidth]{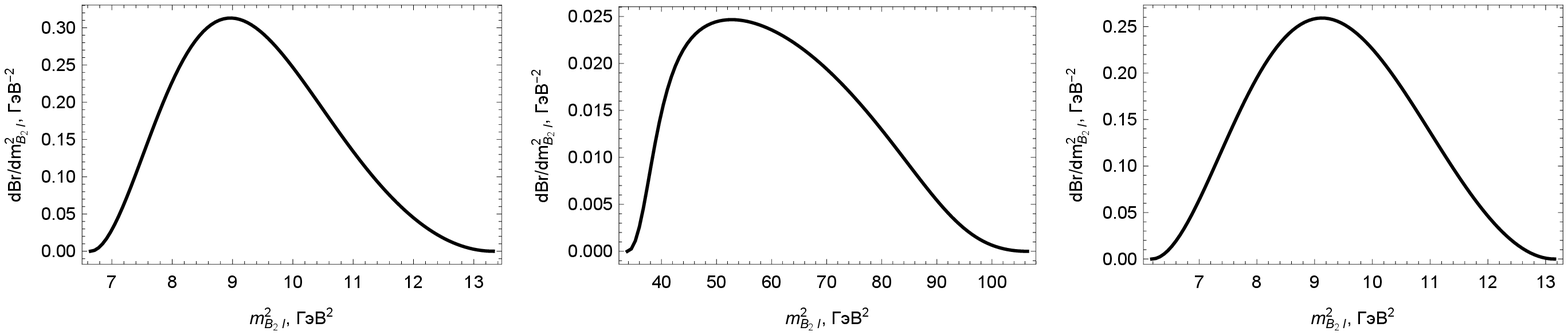}}
\caption{$d\Br/dm_{\B_2 l}^2$ for decays  $\Omega_{cc}^{+} \to \Xi_{c}^{\prime 0} l \nu_l, \Xi_{bb}^{-} \to \Sigma_{b}^{0} l \nu_l, \Xi_{cc}^{+} \to \Xi_{c}^{+} l \nu_l$}
\label{pic:lnul2}
\end{figure}

As for distributions over other Dalitz variables, it is clear that they cannot be obtained in the framework of spectral function formalism. In the case of three-body semileptonic decay, however, we can use a well known Dalitz relation to calculate it. In figure \ref{pic:lnul2}, for example, we show the distributions over the invariant mass of baryon-lepton pair for the same decays.

\begin{table}
\begin{center}
\begin{tabular}{|l|c|c|c|c|c|c|c|c|}
\hline 
\multirow{2}{*}{$\B_1\to \B_2$}& \multicolumn{8}{c|}{Modes}\\
\cline{2-9}
& $l\nu_l$ & $\pi$ & $\rho$ & $2\pi$ & $3\pi$ & $4\pi (ex.\  \omega)$ & $4\pi$ & $5\pi$\\
\hline 

\hline 
$\Xi_{cc}^{++}\to \Lambda_{c}^{+}$&$0.494$&$0.377$&$0.993$&$0.863$&$0.208$&$0.0302$&$0.0637$&$2.2\times 10^{-4}$\\
\hline 
$\Xi_{cc}^{++}\to \Sigma_{c}^{+}$&$0.45$&$0.302$&$1.05$&$0.9$&$0.154$&$0.0101$&$0.0189$&$4.\times 10^{-5}$\\
\hline 
$\Xi_{cc}^{++}\to \Xi_{c}^{+}$&$4.99$&$8.3$&$12.3$&$10.$&$0.634$&$0.0253$&$0.046$&$6.\times 10^{-5}$\\
\hline 
$\Xi_{cc}^{++}\to \Xi_{c}^{\prime+}$&$5.98$&$6.93$&$17.6$&$14.1$&$0.725$&$0.0172$&$0.0307$&$3.\times 10^{-5}$\\
\hline 
$\Xi_{cc}^{+}\to \Sigma_{c}^{0}$&$0.299$&$0.201$&$0.698$&$0.598$&$0.101$&$0.00659$&$0.0124$&$3.\times 10^{-5}$\\
\hline 
$\Xi_{cc}^{+}\to \Xi_{c}^{0}$&$1.65$&$2.75$&$4.08$&$3.33$&$0.21$&$0.00837$&$0.0152$&$2.\times 10^{-5}$\\
\hline 
$\Xi_{cc}^{+}\to \Xi_{c}^{\prime0}$&$1.98$&$2.31$&$5.86$&$4.68$&$0.238$&$0.0056$&$0.00999$&$1.\times 10^{-5}$\\
\hline 
$\Omega_{cc}^{+}\to \Xi_{c}^{0}$&$0.208$&$0.293$&$0.512$&$0.421$&$0.0342$&$0.00166$&$0.00306$&$1.\times 10^{-5}$\\
\hline 
$\Omega_{cc}^{+}\to \Xi_{c}^{\prime0}$&$0.249$&$0.244$&$0.711$&$0.577$&$0.039$&$0.0011$&$0.00199$&$1.\times 10^{-5}$\\
\hline 
$\Omega_{cc}^{+}\to \Omega_{c}^{0}$&$6.66$&$11.2$&$22.6$&$16.9$&$0.265$&$0.00357$&$0.00479$&$1.\times 10^{-5}$\\
\hline 

\end{tabular}
\caption{Branching fractions of $cc$ baryons in percents}
\label{tab:resCC}
\end{center}
\end{table}

\begin{table}
\begin{center}
\begin{tabular}{|l|c|c|c|c|c|c|c|c|}
\hline 
\multirow{2}{*}{$\B_1\to \B_2$}& \multicolumn{8}{c|}{Modes}\\
\cline{2-9}
& $l\nu_l$ & $\pi$ & $\rho$ & $2\pi$ & $3\pi$ & $4\pi (ex. \ \omega)$ & $4\pi$ & $5\pi$\\
\hline 
\hline 
$\Xi_{bc}^{+}\to \Lambda_{b}^{0}$&$0.223$&$0.0438$&$0.486$&$0.415$&$0.0781$&$0.00879$&$0.0187$&$5.\times 10^{-5}$\\
\hline 
$\Xi_{bc}^{+}\to \Sigma_{b}^{0}$&$0.148$&$0.0306$&$0.403$&$0.332$&$0.0292$&$0.00103$&$0.00188$&$1.\times 10^{-5}$\\
\hline 
$\Xi_{bc}^{+}\to \Xi_{b}^{0}$&$2.3$&$0.927$&$5.84$&$4.72$&$0.248$&$0.00837$&$0.0151$&$2.\times 10^{-5}$\\
\hline 
$\Xi_{bc}^{0}\to \Sigma_{b}^{-}$&$0.112$&$0.0233$&$0.306$&$0.251$&$0.0217$&$7.5\times 10^{-4}$&$0.00137$&$1.\times 10^{-5}$\\
\hline 
$\Xi_{bc}^{0}\to \Xi_{b}^{-}$&$0.868$&$0.353$&$2.21$&$1.78$&$0.0922$&$0.00306$&$0.00552$&$1.\times 10^{-5}$\\
\hline 
$\Omega_{bc}^{0}\to \Xi_{b}^{-}$&$0.254$&$0.0384$&$0.511$&$0.443$&$0.105$&$0.0148$&$0.0313$&$1.1\times 10^{-4}$\\
\hline 
$\Omega_{bc}^{0}\to \Omega_{b}^{-}$&$6.03$&$1.25$&$16.6$&$13.6$&$1.07$&$0.0344$&$0.0628$&$7.\times 10^{-5}$\\
\hline 
$\Xi_{bc}^{+}\to \Sigma_{c}^{++}$&$0.0035$&$2.63\times 10^{-6}$&$2.37\times 10^{-4}$&$2.4\times 10^{-4}$&$2.1\times 10^{-4}$&$2.7\times 10^{-4}$&$3.3\times 10^{-4}$&$0.0016$\\
\hline 
$\Xi_{bc}^{+}\to \Xi_{cc}^{++}$&$1.58$&$0.00803$&$0.438$&$0.414$&$0.289$&$0.268$&$0.341$&$0.0703$\\
\hline 
$\Xi_{bc}^{0}\to \Lambda_{c}^{+}$&$3.\times 10^{-4}$&$5.1\times 10^{-7}$&$4.3\times 10^{-5}$&$5.\times 10^{-5}$&$4.\times 10^{-5}$&$4.\times 10^{-5}$&$4.\times 10^{-5}$&$1.6\times 10^{-4}$\\
\hline 
$\Xi_{bc}^{0}\to \Sigma_{c}^{+}$&$7.\times 10^{-4}$&$5.\times 10^{-7}$&$4.6\times 10^{-5}$&$5.\times 10^{-5}$&$4.\times 10^{-5}$&$6.\times 10^{-5}$&$7.\times 10^{-5}$&$3.1\times 10^{-4}$\\
\hline 
$\Xi_{bc}^{0}\to \Xi_{cc}^{+}$&$0.603$&$0.00305$&$0.167$&$0.157$&$0.11$&$0.102$&$0.13$&$0.0267$\\
\hline 
$\Omega_{bc}^{0}\to \Xi_{c}^{+}$&$5.\times 10^{-4}$&$9.4\times 10^{-7}$&$8.8\times 10^{-5}$&$9.\times 10^{-5}$&$7.\times 10^{-5}$&$7.\times 10^{-5}$&$9.\times 10^{-5}$&$2.6\times 10^{-4}$\\
\hline 
$\Omega_{bc}^{0}\to \Omega_{cc}^{+}$&$1.87$&$0.00663$&$0.429$&$0.407$&$0.289$&$0.279$&$0.353$&$0.0784$\\
\hline 
\end{tabular}
\caption{Branching fractions of $bc$ baryons in percents}
\label{tab:resBC}
\end{center}
\end{table}

\begin{table}
\begin{center}
\begin{tabular}{|l|c|c|c|c|c|c|c|c|}
\hline 
\multirow{2}{*}{$\B_1\to \B_2$}& \multicolumn{8}{c|}{Modes}\\
\cline{2-9}
& $l\nu_l$ & $\pi$ & $\rho$ & $2\pi$ & $3\pi$ & $4\pi (ex. \ \omega)$ & $4\pi$ & $5\pi$\\
\hline 
\hline 
$\Xi_{bb}^{0}\to \Sigma_{b}^{+}$&$0.0043$&$1.43\times 10^{-6}$&$4.34\times 10^{-4}$&$4.3\times 10^{-4}$&$3.8\times 10^{-4}$&$4.7\times 10^{-4}$&$5.8\times 10^{-4}$&$0.00128$\\
\hline 
$\Xi_{bb}^{0}\to \Xi_{bc}^{+}$&$2.59$&$0.00347$&$0.568$&$0.54$&$0.399$&$0.397$&$0.499$&$0.113$\\
\hline 
$\Xi_{bb}^{0}\to \Xi_{bc}^{\prime+}$&$1.15$&$6.26\times 10^{-4}$&$0.128$&$0.125$&$0.113$&$0.141$&$0.172$&$0.0499$\\
\hline 
$\Xi_{bb}^{-}\to \Lambda_{b}^{0}$&$0.0011$&$8.3\times 10^{-7}$&$2.23\times 10^{-4}$&$2.2\times 10^{-4}$&$1.7\times 10^{-4}$&$1.6\times 10^{-4}$&$2.1\times 10^{-4}$&$3.1\times 10^{-4}$\\
\hline 
$\Xi_{bb}^{-}\to \Sigma_{b}^{0}$&$0.0022$&$7.2\times 10^{-7}$&$2.18\times 10^{-4}$&$2.2\times 10^{-4}$&$1.9\times 10^{-4}$&$2.4\times 10^{-4}$&$2.9\times 10^{-4}$&$6.5\times 10^{-4}$\\
\hline 
$\Xi_{bb}^{-}\to \Xi_{bc}^{0}$&$2.62$&$0.00349$&$0.572$&$0.544$&$0.402$&$0.4$&$0.504$&$0.115$\\
\hline 
$\Xi_{bb}^{-}\to \Xi_{bc}^{\prime0}$&$1.16$&$6.23\times 10^{-4}$&$0.128$&$0.125$&$0.113$&$0.142$&$0.172$&$0.05$\\
\hline 
$\Omega_{bb}^{-}\to \Xi_{b}^{0}$&$0.002$&$1.64\times 10^{-6}$&$4.51\times 10^{-4}$&$4.4\times 10^{-4}$&$3.3\times 10^{-4}$&$3.2\times 10^{-4}$&$4.1\times 10^{-4}$&$4.7\times 10^{-4}$\\
\hline 
$\Omega_{bb}^{-}\to \Xi_{b}^{\prime0}$&$0.0044$&$1.44\times 10^{-6}$&$4.53\times 10^{-4}$&$4.5\times 10^{-4}$&$4.\times 10^{-4}$&$4.9\times 10^{-4}$&$6.\times 10^{-4}$&$0.00134$\\
\hline 
$\Omega_{bb}^{-}\to \Omega_{bc}^{0}$&$4.81$&$0.00702$&$1.13$&$1.08$&$0.792$&$0.776$&$0.979$&$0.216$\\
\hline 
$\Omega_{bb}^{-}\to \Omega_{bc}^{\prime0}$&$2.13$&$0.00126$&$0.256$&$0.251$&$0.226$&$0.28$&$0.342$&$0.0966$\\
\hline 
\end{tabular}
\caption{Branching fractions of $bb$ baryons in percents}
\label{tab:resBB}
\end{center}
\end{table}

\renewcommand{\arraystretch}{1.4}

\section{Conclusion}
\label{sec:conclusion}

In the presented paper we have considered exclusive decays of the ground states of doubly heavy baryons $\Xi_{Q_{1}Q_{2}}$, $\Omega_{Q_{1}Q_{2}}$ (where $Q_{1,2}=c, b$) with production of the leptonic pair or the system of charged $\pi$ mesons. According to the factorization theorem the widths  of  these processes can be written as convolution of semileptonic decays and spectral functions, that are connected with virtual $W$-boson transition into final system of light particles. First stage can be described in terms of form factors of the weak decay, that can be calculated, for example, in frame of potential models. The spectral functions, on the other hand, can be calculated, for example, from analysis of $\tau$ lepton decays.  

In our article we gave analytical expressions for distributions of decay width over squared transferred momentum. Using known parametrization of form factors of doubly heavy baryons and spectral functions we gave numerical predictions of these distributions and integrated branching fractions. According to our results, for some of the processes branchings are quite large, that can lead to their experimental observation. It should be noted, that semileptonic decays of doubly heavy baryons were considered also in a number of other works (see, for example, \cite{PerezMarcial:1989yh, Albertus:2012nd, Shi:2019hbf}). The results of these works have some difference with ours, which is caused by the difference in form-factors' parametrization. Thereby, we think it is extremely interesting to try to experimentally observe marked decays.

In future we plan to continue our work in this area. For example, it would be interesting to study the production  of light mesons in decays of excited $P$-waved states of doubly heavy baryons.
In addition, we plan to creation of based on the presented theoretical models Monte-Carlo generators, that are required for comparison of theoretical predictions with forthcoming experimental results.

The authors would like  grateful to A. K. Likhoded for productive discussions and help in preparation of this article.

%

\appendix

\section{Parametrization of form factors}\label{app2}
As it was shown in section II, using results \cite{Wang:2017mqp} in case of $\B_1\to \B_2W$ transition we can obtain dependence of form factors on squared transferred momentum for scalar and axial diquarks. In this paper it is more convenient to parametrize not the form factors themselves, but their sum
\begin{align}
F(q^{2}) &= c_S F_S(q^{2}) + c_A F_A(q^{2})
\end{align}
In our article we use the following parametrization:
\begin{align}
F(q^2) &= F(0)(1+\alpha q^2+\beta q^4 + \gamma q^6),
\end{align}
where parameters $F(0)$, $\alpha$, $\beta$, and $\gamma$ for different initial and final states are given in tables \ref{tab:ffCC}-\ref{tab:ggBB}.

\begin{table}
\begin{center}
\begin{tabular}{|l||c|c|c|c||c|c|c|c|}
\hline 
\multirow{2}{*}{$\B_1\to \B_2$}& \multicolumn{4}{c||}{$f_1$} & \multicolumn{4}{c|}{$f_2$}\\
\cline{2-5}\cline{6-9}
& $F(0)$ & $\alpha, \GeV^{-2}$ & $\beta, \GeV^{-4}$ & $\gamma, \GeV^{-6}$ & $F(0)$ & $\alpha, \GeV^{-2}$ & $\beta,\GeV^{-4}$ & $\gamma, \GeV^{-6}$\\
\hline 
\hline 
$\Xi_{cc}^{++}\to \Lambda_{c}^{+}$&$0.791$&$0.386$&$0.118$&$0.016$&$-0.00794$&$-0.481$&$-0.405$&$-0.2$\\
\hline 
$\Xi_{cc}^{++}\to \Sigma_{c}^{+}$&$-0.467$&$0.294$&$0.0331$&$0.0417$&$1.04$&$0.418$&$0.108$&$0.037$\\
\hline 
$\Xi_{cc}^{++}\to \Xi_{c}^{+}$&$0.914$&$0.348$&$0.0818$&$0.0187$&$0.0116$&$1.31$&$0.513$&$0.15$\\
\hline 
$\Xi_{cc}^{++}\to \Xi_{c}^{\prime+}$&$-0.538$&$0.247$&$0.0384$&$0.0213$&$1.11$&$0.366$&$0.0863$&$0.0276$\\
\hline 
$\Xi_{cc}^{+}\to \Sigma_{c}^{0}$&$-0.661$&$0.294$&$0.0332$&$0.0416$&$1.47$&$0.418$&$0.108$&$0.037$\\
\hline 
$\Xi_{cc}^{+}\to \Xi_{c}^{0}$&$0.914$&$0.348$&$0.0818$&$0.0187$&$0.0116$&$1.31$&$0.513$&$0.15$\\
\hline 
$\Xi_{cc}^{+}\to \Xi_{c}^{\prime0}$&$-0.538$&$0.247$&$0.0384$&$0.0213$&$1.11$&$0.366$&$0.0863$&$0.0276$\\
\hline 
$\Omega_{cc}^{+}\to \Xi_{c}^{0}$&$-0.783$&$0.406$&$0.117$&$0.0191$&$0.0214$&$0.194$&$-0.0127$&$-0.0214$\\
\hline 
$\Omega_{cc}^{+}\to \Xi_{c}^{\prime0}$&$-0.462$&$0.308$&$0.0495$&$0.0408$&$1.05$&$0.425$&$0.116$&$0.0334$\\
\hline 
$\Omega_{cc}^{+}\to \Omega_{c}^{0}$&$-0.754$&$0.263$&$0.047$&$0.0205$&$1.59$&$0.376$&$0.0926$&$0.0244$\\
\hline 
\end{tabular}
\caption{Parameters of form factors $f_1$, $f_2$ for $cc$ baryons}
\label{tab:ffCC}
\end{center}
\end{table}

\begin{table}
\begin{center}
\begin{tabular}{|l||c|c|c|c||c|c|c|c|}
\hline 
\multirow{2}{*}{$\B_1\to \B_2$}& \multicolumn{4}{c||}{$g_1$} & \multicolumn{4}{c|}{$g_2$}\\
\cline{2-5}\cline{6-9}
& $F(0)$ & $\alpha, \GeV^{-2}$ & $\beta, \GeV^{-4}$ & $\gamma, \GeV^{-6}$ & $F(0)$ & $\alpha, \GeV^{-2}$ & $\beta,\GeV^{-4}$ & $\gamma, \GeV^{-6}$\\
\hline 
\hline 
$\Xi_{cc}^{++}\to \Lambda_{c}^{+}$&$0.224$&$0.235$&$0.0386$&$-0.0045$&$-0.0482$&$0.845$&$-1.14$&$0.295$\\
\hline 
$\Xi_{cc}^{++}\to \Sigma_{c}^{+}$&$-0.624$&$0.244$&$0.0378$&$0.00399$&$0.0447$&$1.61$&$-1.68$&$0.321$\\
\hline 
$\Xi_{cc}^{++}\to \Xi_{c}^{+}$&$0.258$&$0.208$&$0.0262$&$-2.8\times 10^{-4}$&$-0.0608$&$0.364$&$0.289$&$-0.16$\\
\hline 
$\Xi_{cc}^{++}\to \Xi_{c}^{\prime+}$&$-0.728$&$0.216$&$0.0305$&$0.00386$&$0.0783$&$0.649$&$0.331$&$-0.187$\\
\hline 
$\Xi_{cc}^{+}\to \Sigma_{c}^{0}$&$-0.883$&$0.244$&$0.0378$&$0.00399$&$0.0632$&$1.61$&$-1.68$&$0.319$\\
\hline 
$\Xi_{cc}^{+}\to \Xi_{c}^{0}$&$0.258$&$0.208$&$0.0262$&$-2.8\times 10^{-4}$&$-0.0608$&$0.364$&$0.289$&$-0.16$\\
\hline 
$\Xi_{cc}^{+}\to \Xi_{c}^{\prime0}$&$-0.728$&$0.216$&$0.0305$&$0.00386$&$0.0783$&$0.649$&$0.33$&$-0.187$\\
\hline 
$\Omega_{cc}^{+}\to \Xi_{c}^{0}$&$-0.222$&$0.249$&$0.0375$&$-0.00273$&$0.0535$&$0.733$&$-0.778$&$0.103$\\
\hline 
$\Omega_{cc}^{+}\to \Xi_{c}^{\prime0}$&$-0.618$&$0.253$&$0.0398$&$0.00308$&$0.0511$&$1.3$&$-0.874$&$-0.133$\\
\hline 
$\Omega_{cc}^{+}\to \Omega_{c}^{0}$&$-1.02$&$0.225$&$0.0329$&$0.00381$&$0.119$&$0.671$&$0.297$&$-0.159$\\
\hline 
\end{tabular}
\caption{Parameters of form factors $g_1$, $g_2$ for $cc$ baryons}
\label{tab:ggCC}
\end{center}
\end{table}

\begin{table}
\begin{center}
\begin{tabular}{|l||c|c|c|c||c|c|c|c|}
\hline 
\multirow{2}{*}{$\B_1\to \B_2$}& \multicolumn{4}{c||}{$f_1$} & \multicolumn{4}{c|}{$f_2$}\\
\cline{2-5}\cline{6-9}
& $F(0)$ & $\alpha, \GeV^{-2}$ & $\beta, \GeV^{-4}$ & $\gamma, \GeV^{-6}$ & $F(0)$ & $\alpha, \GeV^{-2}$ & $\beta,\GeV^{-4}$ & $\gamma, \GeV^{-6}$\\
\hline 
\hline 
$\Xi_{bc}^{+}\to \Lambda_{b}^{0}$&$0.554$&$0.421$&$0.205$&$-0.0498$&$-0.297$&$0.425$&$0.211$&$-0.05$\\
\hline 
$\Xi_{bc}^{+}\to \Sigma_{b}^{0}$&$-0.32$&$0.406$&$0.107$&$0.0123$&$1.54$&$0.453$&$0.155$&$-0.0203$\\
\hline 
$\Xi_{bc}^{+}\to \Xi_{b}^{0}$&$0.627$&$0.405$&$0.118$&$-0.00891$&$-0.301$&$0.423$&$0.131$&$-0.0101$\\
\hline 
$\Xi_{bc}^{0}\to \Sigma_{b}^{-}$&$-0.453$&$0.407$&$0.107$&$0.0123$&$2.17$&$0.453$&$0.155$&$-0.0201$\\
\hline 
$\Xi_{bc}^{0}\to \Xi_{b}^{-}$&$0.627$&$0.405$&$0.118$&$-0.00877$&$-0.301$&$0.423$&$0.13$&$-0.00994$\\
\hline 
$\Omega_{bc}^{0}\to \Xi_{b}^{-}$&$-0.554$&$0.41$&$0.23$&$-0.0655$&$0.313$&$0.416$&$0.263$&$-0.0857$\\
\hline 
$\Omega_{bc}^{0}\to \Omega_{b}^{-}$&$-0.512$&$0.366$&$0.0841$&$0.013$&$2.38$&$0.416$&$0.121$&$-0.00663$\\
\hline 
$\Xi_{bc}^{+}\to \Sigma_{c}^{++}$&$-0.094$&$0.118$&$0.0027$&$-3.91\times 10^{-4}$&$0.124$&$0.232$&$-0.0196$&$3.63\times 10^{-4}$\\
\hline 
$\Xi_{bc}^{+}\to \Xi_{cc}^{++}$&$0.771$&$0.0531$&$0.00247$&$-1.02\times 10^{-4}$&$-0.0579$&$0.0459$&$3.56\times 10^{-4}$&$1.02\times 10^{-4}$\\
\hline 
$\Xi_{bc}^{0}\to \Lambda_{c}^{+}$&$-0.104$&$0.194$&$-0.0165$&$3.17\times 10^{-4}$&$-0.0262$&$0.265$&$-0.0254$&$5.63\times 10^{-4}$\\
\hline 
$\Xi_{bc}^{0}\to \Sigma_{c}^{+}$&$-0.0664$&$0.118$&$0.00268$&$-3.9\times 10^{-4}$&$0.0875$&$0.232$&$-0.0196$&$3.64\times 10^{-4}$\\
\hline 
$\Xi_{bc}^{0}\to \Xi_{cc}^{+}$&$0.771$&$0.0531$&$0.00247$&$-1.02\times 10^{-4}$&$-0.0579$&$0.0459$&$3.56\times 10^{-4}$&$1.02\times 10^{-4}$\\
\hline 
$\Omega_{bc}^{0}\to \Xi_{c}^{+}$&$-0.0944$&$0.203$&$-0.0191$&$4.09\times 10^{-4}$&$-0.0239$&$0.26$&$-0.0269$&$6.38\times 10^{-4}$\\
\hline 
$\Omega_{bc}^{0}\to \Omega_{cc}^{+}$&$0.745$&$0.0544$&$0.003$&$-1.52\times 10^{-4}$&$-0.0669$&$0.0497$&$0.00135$&$3.14\times 10^{-5}$\\
\hline 
\end{tabular}
\caption{Parameters of form factors $f_1$, $f_2$ for $bc$ baryons}
\label{tab:ffBC}
\end{center}
\end{table}

\begin{table}
\begin{center}
\begin{tabular}{|l||c|c|c|c||c|c|c|c|}
\hline 
\multirow{2}{*}{$\B_1\to \B_2$}& \multicolumn{4}{c||}{$g_1$} & \multicolumn{4}{c|}{$g_2$}\\
\cline{2-5}\cline{6-9}
& $F(0)$ & $\alpha, \GeV^{-2}$ & $\beta, \GeV^{-4}$ & $\gamma, \GeV^{-6}$ & $F(0)$ & $\alpha, \GeV^{-2}$ & $\beta,\GeV^{-4}$ & $\gamma, \GeV^{-6}$\\
\hline 
\hline 
$\Xi_{bc}^{+}\to \Lambda_{b}^{0}$&$0.147$&$0.295$&$0.0589$&$-0.0201$&$-0.095$&$1.32$&$-0.664$&$-0.0546$\\
\hline 
$\Xi_{bc}^{+}\to \Sigma_{b}^{0}$&$-0.414$&$0.292$&$0.0462$&$-0.0124$&$0.174$&$0.816$&$0.194$&$-0.479$\\
\hline 
$\Xi_{bc}^{+}\to \Xi_{b}^{0}$&$0.167$&$0.275$&$0.042$&$-0.00782$&$-0.11$&$1.11$&$-0.251$&$-0.0448$\\
\hline 
$\Xi_{bc}^{0}\to \Sigma_{b}^{-}$&$-0.586$&$0.292$&$0.0461$&$-0.0123$&$0.247$&$0.814$&$0.199$&$-0.482$\\
\hline 
$\Xi_{bc}^{0}\to \Xi_{b}^{-}$&$0.167$&$0.275$&$0.0419$&$-0.00779$&$-0.111$&$1.08$&$-0.225$&$-0.0543$\\
\hline 
$\Omega_{bc}^{0}\to \Xi_{b}^{-}$&$-0.146$&$0.296$&$0.0642$&$-0.0226$&$0.104$&$1.23$&$-0.717$&$0.00824$\\
\hline 
$\Omega_{bc}^{0}\to \Omega_{b}^{-}$&$-0.67$&$0.272$&$0.0407$&$-0.00635$&$0.32$&$0.181$&$1.08$&$-0.654$\\
\hline 
$\Xi_{bc}^{+}\to \Sigma_{c}^{++}$&$-0.139$&$0.123$&$-0.00335$&$-1.2\times 10^{-4}$&$-0.00321$&$0.171$&$0.0308$&$-0.00178$\\
\hline 
$\Xi_{bc}^{+}\to \Xi_{cc}^{++}$&$0.511$&$0.0474$&$0.00162$&$-2.64\times 10^{-5}$&$-0.0669$&$0.057$&$0.00324$&$-1.53\times 10^{-4}$\\
\hline 
$\Xi_{bc}^{0}\to \Lambda_{c}^{+}$&$-0.0428$&$0.166$&$-0.00808$&$6.56\times 10^{-6}$&$0.0194$&$0.17$&$-0.0141$&$2.58\times 10^{-4}$\\
\hline 
$\Xi_{bc}^{0}\to \Sigma_{c}^{+}$&$-0.098$&$0.123$&$-0.00336$&$-1.2\times 10^{-4}$&$-0.00226$&$0.172$&$0.0307$&$-0.00178$\\
\hline 
$\Xi_{bc}^{0}\to \Xi_{cc}^{+}$&$0.511$&$0.0474$&$0.00162$&$-2.64\times 10^{-5}$&$-0.0669$&$0.057$&$0.00324$&$-1.53\times 10^{-4}$\\
\hline 
$\Omega_{bc}^{0}\to \Xi_{c}^{+}$&$-0.0378$&$0.205$&$-0.0143$&$1.97\times 10^{-4}$&$0.0186$&$0.181$&$-0.017$&$3.59\times 10^{-4}$\\
\hline 
$\Omega_{bc}^{0}\to \Omega_{cc}^{+}$&$0.493$&$0.0491$&$0.00209$&$-6.17\times 10^{-5}$&$-0.0713$&$0.0578$&$0.00424$&$-2.55\times 10^{-4}$\\
\hline 
\end{tabular}
\caption{Parameters of form factors $g_1$, $g_2$ for $bc$ baryons}
\label{tab:ggBC}
\end{center}
\end{table}

\begin{table}
\begin{center}
\begin{tabular}{|l||c|c|c|c||c|c|c|c|}
\hline 
\multirow{2}{*}{$\B_1\to \B_2$}& \multicolumn{4}{c||}{$f_1$} & \multicolumn{4}{c|}{$f_2$}\\
\cline{2-5}\cline{6-9}
& $F(0)$ & $\alpha, \GeV^{-2}$ & $\beta, \GeV^{-4}$ & $\gamma, \GeV^{-6}$ & $F(0)$ & $\alpha, \GeV^{-2}$ & $\beta,\GeV^{-4}$ & $\gamma, \GeV^{-6}$\\
\hline 
\hline 
$\Xi_{bb}^{0}\to \Sigma_{b}^{+}$&$-0.0751$&$0.254$&$-0.0267$&$6.45\times 10^{-4}$&$0.207$&$0.194$&$-0.0253$&$7.04\times 10^{-4}$\\
\hline 
$\Xi_{bb}^{0}\to \Xi_{bc}^{+}$&$0.592$&$0.0674$&$0.00469$&$-4.29\times 10^{-4}$&$-0.266$&$0.0701$&$0.00503$&$-5.03\times 10^{-4}$\\
\hline 
$\Xi_{bb}^{0}\to \Xi_{bc}^{\prime+}$&$-0.00381$&$-0.858$&$0.0343$&$0.0124$&$0.527$&$0.0711$&$0.00515$&$-5.22\times 10^{-4}$\\
\hline 
$\Xi_{bb}^{-}\to \Lambda_{b}^{0}$&$-0.101$&$0.168$&$-0.0213$&$5.62\times 10^{-4}$&$6.04\times 10^{-4}$&$0.169$&$-0.0218$&$5.83\times 10^{-4}$\\
\hline 
$\Xi_{bb}^{-}\to \Sigma_{b}^{0}$&$-0.0531$&$0.254$&$-0.0267$&$6.45\times 10^{-4}$&$0.146$&$0.194$&$-0.0253$&$7.03\times 10^{-4}$\\
\hline 
$\Xi_{bb}^{-}\to \Xi_{bc}^{0}$&$0.592$&$0.0674$&$0.00469$&$-4.29\times 10^{-4}$&$-0.266$&$0.0701$&$0.00503$&$-5.03\times 10^{-4}$\\
\hline 
$\Xi_{bb}^{-}\to \Xi_{bc}^{\prime0}$&$-0.00384$&$-0.861$&$0.0361$&$0.0122$&$0.527$&$0.0711$&$0.00514$&$-5.22\times 10^{-4}$\\
\hline 
$\Omega_{bb}^{-}\to \Xi_{b}^{0}$&$-0.098$&$0.165$&$-0.0216$&$5.84\times 10^{-4}$&$0.00121$&$0.167$&$-0.0221$&$6.03\times 10^{-4}$\\
\hline 
$\Omega_{bb}^{-}\to \Xi_{b}^{\prime0}$&$-0.0527$&$0.248$&$-0.0271$&$6.75\times 10^{-4}$&$0.146$&$0.186$&$-0.0248$&$6.96\times 10^{-4}$\\
\hline 
$\Omega_{bb}^{-}\to \Omega_{bc}^{0}$&$0.586$&$0.0689$&$0.00465$&$-4.5\times 10^{-4}$&$-0.27$&$0.0713$&$0.00496$&$-5.23\times 10^{-4}$\\
\hline 
$\Omega_{bb}^{-}\to \Omega_{bc}^{\prime0}$&$-0.00302$&$-1.$&$0.032$&$0.0167$&$0.527$&$0.0723$&$0.00508$&$-5.43\times 10^{-4}$\\
\hline 
\end{tabular}
\caption{Parameters of form factors $f_1$, $f_2$ for $bb$ baryons}
\label{tab:ffBB}
\end{center}
\end{table}

\begin{table}
\begin{center}
\begin{tabular}{|l||c|c|c|c||c|c|c|c|}
\hline 
\multirow{2}{*}{$\B_1\to \B_2$}& \multicolumn{4}{c||}{$g_1$} & \multicolumn{4}{c|}{$g_2$}\\
\cline{2-5}\cline{6-9}
& $F(0)$ & $\alpha, \GeV^{-2}$ & $\beta, \GeV^{-4}$ & $\gamma, \GeV^{-6}$ & $F(0)$ & $\alpha, \GeV^{-2}$ & $\beta,\GeV^{-4}$ & $\gamma, \GeV^{-6}$\\
\hline 
\hline 
$\Xi_{bb}^{0}\to \Sigma_{b}^{+}$&$-0.116$&$0.208$&$-0.022$&$5.26\times 10^{-4}$&$-0.0231$&$0.122$&$-0.015$&$3.78\times 10^{-4}$\\
\hline 
$\Xi_{bb}^{0}\to \Xi_{bc}^{+}$&$0.376$&$0.0606$&$0.00387$&$-3.02\times 10^{-4}$&$-0.0131$&$0.11$&$0.00769$&$-5.88\times 10^{-4}$\\
\hline 
$\Xi_{bb}^{0}\to \Xi_{bc}^{\prime+}$&$-0.312$&$0.0595$&$0.00368$&$-2.75\times 10^{-4}$&$-0.0448$&$0.0527$&$0.00293$&$-2.69\times 10^{-4}$\\
\hline 
$\Xi_{bb}^{-}\to \Lambda_{b}^{0}$&$-0.0336$&$0.204$&$-0.0225$&$5.51\times 10^{-4}$&$0.0112$&$0.243$&$-0.026$&$6.28\times 10^{-4}$\\
\hline 
$\Xi_{bb}^{-}\to \Sigma_{b}^{0}$&$-0.0822$&$0.208$&$-0.022$&$5.26\times 10^{-4}$&$-0.0163$&$0.122$&$-0.015$&$3.77\times 10^{-4}$\\
\hline 
$\Xi_{bb}^{-}\to \Xi_{bc}^{0}$&$0.376$&$0.0606$&$0.00388$&$-3.03\times 10^{-4}$&$-0.0135$&$0.109$&$0.00759$&$-5.81\times 10^{-4}$\\
\hline 
$\Xi_{bb}^{-}\to \Xi_{bc}^{\prime0}$&$-0.312$&$0.0594$&$0.00368$&$-2.76\times 10^{-4}$&$-0.0454$&$0.0528$&$0.00294$&$-2.7\times 10^{-4}$\\
\hline 
$\Omega_{bb}^{-}\to \Xi_{b}^{0}$&$-0.0331$&$0.204$&$-0.0231$&$5.75\times 10^{-4}$&$0.0107$&$0.245$&$-0.0269$&$6.67\times 10^{-4}$\\
\hline 
$\Omega_{bb}^{-}\to \Xi_{b}^{\prime0}$&$-0.0801$&$0.208$&$-0.0226$&$5.5\times 10^{-4}$&$-0.0161$&$0.117$&$-0.0149$&$3.84\times 10^{-4}$\\
\hline 
$\Omega_{bb}^{-}\to \Omega_{bc}^{0}$&$0.371$&$0.062$&$0.00389$&$-3.09\times 10^{-4}$&$-0.0152$&$0.111$&$0.00592$&$-5.31\times 10^{-4}$\\
\hline 
$\Omega_{bb}^{-}\to \Omega_{bc}^{\prime0}$&$-0.308$&$0.0608$&$0.00368$&$-2.8\times 10^{-4}$&$-0.0436$&$0.0523$&$0.00304$&$-2.82\times 10^{-4}$\\
\hline 
\end{tabular}
\caption{Parameters of form factors $g_1$, $g_2$ for $bb$ baryons}
\label{tab:ggBB}
\end{center}
\end{table}

\end{document}